# Empirical non-parametric estimation of the Fisher Information

Visar Berisha* and Alfred O. Hero†

**Abstract**

The Fisher information matrix (FIM) is a foundational concept in statistical signal processing. The FIM depends on the probability distribution, assumed to belong to a smooth parametric family. Traditional approaches to estimating the FIM require estimating the probability distribution function (PDF), or its parameters, along with its gradient or Hessian. However, in many practical situations the PDF of the data is not known but the statistician has access to an observation sample for any parameter value. Here we propose a method of estimating the FIM directly from sampled data that does not require knowledge of the underlying PDF. The method is based on non-parametric estimation of an $f$-divergence over a local neighborhood of the parameter space and a relation between curvature of the $f$-divergence and the FIM. Thus we obtain an empirical estimator of the FIM that does not require density estimation and is asymptotically consistent. We empirically evaluate the validity of our approach using two experiments.

## 1 Introduction

The Fisher information matrix (FIM) is an important quantity in signal processing and statistical estimation. It can be used to benchmark the performance of an estimator (via the Cramér-Rao Lower Bound (CRLB)) [1], design an optimal experiment to maximize the trace or determinant of the FIM [2], or develop super-efficient estimators [3]. The FIM can be computed as the covariance matrix of the gradient of the log likelihood function (the score function). However, the statistical model is often unknown or the covariance computation may be intractable to manipulate. In some cases one has access to a generative model, a black box that can generate multiple realizations from a distribution for various settings of its parameters, e.g., when one can generate data from a controlled experiment with accessible and tunable operating parameters [4, 5, 6]. In such cases it makes sense to try and compute the FIM empirically from multiple perturbed experiments. This is the case addressed here.

*V. Berisha is with the School of Electrical, Computer, and Energy Engineering and the Department of Speech and Hearing Science, Arizona State University, e-mail: visar@asu.edu. This author gratefully acknowledges partial support of this work by ONR grant N000141410722.

†A. Hero is with the Department of Electrical Engineering and Computer Science, University of Michigan, e-mail: hero@eecs.umich.edu. This author gratefully acknowledges partial support of this work by ARO grant W911NF-11-1-0391 and NSF grant CCF-1217880.

The first approach that might come to mind would be to perform non-parametric density estimation for each of the perturbed experiments and then compute a numerical gradient of the log density followed by sample averaging. Another, related, approach would be to estimate an $f$-divergence measure between the estimated densities and use the relation between the $f$-divergence and the FIM through its second variational term in the Taylor expansion. However, these density estimate-and-plug strategies are not the most natural approaches. Indeed it seems ill-advised to approach the problem of estimation of a finite dimensional quantity, e.g. the elements of the FIM, by first generating an estimate of an infinite dimensional quantity, e.g., the density function. Here we introduce a more direct way of estimating the FIM that estimates the $f$-divergence measure directly from the data, and does not require density estimation. We empirically evaluate this procedure on two experiments: (1) comparing the non-parametric estimate of the FIM to an estimate based on a closed-form expression for the FIM in a setting where the closed-form expression is available; and (2) using the FIM as a proxy for the difficulty of an estimation problem (based on its relationship with the CRLB) using data from a cochlear implant simulator.

## 2 A Non-Parametric $f$-Divergence

Given two probability density functions $p$ and $q$ with domain $\mathbb{R}^K$ and a parameter $\alpha \in (0\ldots 1)$, let us consider the following divergence between the two distributions:

$$D_\alpha(p,q) = \frac{1}{4\alpha(1-\alpha)} \left[ \int \frac{(\alpha p(\mathbf{x}) - (1-\alpha)q(\mathbf{x}))^2}{\alpha p(\mathbf{x}) + (1-\alpha)q(\mathbf{x})} d\mathbf{x} - (2\alpha - 1)^2 \right] \quad (1)$$

The divergence measure in (1) belongs to the class of $f$-divergences or Ali-Silvey distances [7]. Intuitively, an $f$-divergence is an average of the ratio of two distributions, weighted by some function $f(t)$: $D_f(p,q) = \int f(\frac{p(\mathbf{x})}{q(\mathbf{x})}) q(\mathbf{x}) d\mathbf{x}$. Many common divergences used in statistical signal processing fall in this category, including the KL-divergence, the Hellinger distance, the total variation distance, etc. [7]. For $D_\alpha(p,q)$, the corresponding function $f(t)$ is,

$$f(t) = \frac{1}{4\alpha(1-\alpha)} \left[ \frac{(\alpha t - (1-\alpha))^2}{\alpha t + (1-\alpha)} - (2\alpha - 1)^2 \right]. \quad (2)$$

Furthermore, $f(t)$ is defined for all $t > 0$, is convex, and $f(1) = 0$, consistent with the requirements of the definition of an $f$-divergence [7]. Indeed, for the special case of $\alpha = \frac{1}{2}$, the divergence in (1) becomes the symmetric $\chi^2$ $f$-divergence in [8] and is similar to the Rukhin $f$-divergence in [9].

The divergence in (1) has the remarkable property that it can be estimated directly without estimation or plug-in of the densities $p$ and $q$ based on an extension of the Friedman-Rafsky (FR) multi-variate two sample test statistic [10]. Let us consider sample realizations from $p$ and $q$, denoted by $\mathbf{X}_p \in \mathbb{R}^{N_p \times K}$, $\mathbf{X}_q \in \mathbb{R}^{N_q \times K}$, and $\alpha = \frac{N_p}{N_p + N_q}$. The FR test statistic, $\mathcal{C}(\mathbf{X}_p, \mathbf{X}_q)$, is constructed by first generating a Euclidean minimal spanning tree (MST) on the concatenated data set, $\mathbf{X}_p \cup \mathbf{X}_q$, and then counting the number of edges connecting a data point from $p$ to a data point from $q$. The test assumes a unique MST for $\mathbf{X}_p \cup \mathbf{X}_q$ - therefore all inter point distances between data points must be distinct. However, this assumption is not restrictive since the MST is unique with probability one when $p$ and $q$ are Lebesgue continuous densities. Henze and Penrose proved the following theorem related to this test statistic [11]:

**Theorem 1.** *As $N_p \to \infty$ and $N_q \to \infty$ in a linked manner such that $\frac{N_p}{N_p + N_q} \to \alpha$,*

$$\frac{\mathcal{C}(\mathbf{X}_p, \mathbf{X}_q)}{N_p + N_q} \to A_\alpha(p, q) \quad \text{almost surely},$$

*where $A_\alpha(p, q) = 2\alpha(1-\alpha) \int \frac{p(\mathbf{x})q(\mathbf{x})}{\alpha p(\mathbf{x}) + (1-\alpha)q(\mathbf{x})} d\mathbf{x}$.*

Combining the results of Theorem 1 with the definition of the divergence function in (1) we see that

$$D_\alpha(p, q) = 1 - \frac{A_\alpha(p, q)}{2\alpha(1-\alpha)}. \tag{3}$$

Further, under the same conditions as those in Theorem 1,

$$1 - \mathcal{C}(\mathbf{X}_p, \mathbf{X}_q) \frac{N_p + N_q}{2N_p N_q} \to D_\alpha(p, q). \tag{4}$$

It is important to note that the parameter $\alpha$ does not directly appear in (4) but rather is controlled by the ratio of samples from $p(\mathbf{x})$ to the total number of samples - it is interpreted as a prior probability that a sample comes from $p$.

The left side of (4) specifies an empirical estimator of the $f$-divergence in (1) that does not require estimation of the densities $p$ and $q$. We use this fact to derive a non-parametric estimator of the Fisher Information directly from data.

## 3 FIM Estimation

Let the densities $p$ and $q$ with domain $\mathbb{R}^K$ be parameterized by a multidimensional parameter $\boldsymbol{\theta}$ with domain $\mathbb{R}^d$ and consider the case where $q$ is a directional perturbation of $p$ about some particular value of $\boldsymbol{\theta}$:

$$q = p_{\boldsymbol{\theta} + \mathbf{u}}, \tag{5}$$

where $\boldsymbol{\theta} + \mathbf{u}$ is a small perturbation around $\boldsymbol{\theta}$. Amari and Cichocki showed that *any* $f$-divergence induces a unique information monotonic Riemannian metric, given by the $d \times d$ Fisher information matrix, $\mathbf{F}_{\boldsymbol{\theta}} = E_{\boldsymbol{\theta}}[\nabla p_{\boldsymbol{\theta}} \nabla p_{\boldsymbol{\theta}}^T]$ (Thm. 5 in [12]). The FIM is a symmetric positive semidefinite definite matrix with $d(d+2)/2$ distinct entries. Using the Taylor expansion they show that any $f$-divergence, $D_f(p_{\boldsymbol{\theta}}, p_{\boldsymbol{\theta}+\mathbf{u}})$, is related to the FIM through the asymptotic relation

$$D_f(p_{\boldsymbol{\theta}}, p_{\boldsymbol{\theta}+\mathbf{u}}) = \frac{1}{2}\mathbf{u}^T \mathbf{F}_{\boldsymbol{\theta}} \mathbf{u} + o(\|\mathbf{u}\|^2). \tag{6}$$

Combining this relationship with the non-parametric $f$-divergence from the previous section, we define

$$Q_{pq} = 2D_\alpha(p, q) \approx \mathbf{u}^T \mathbf{F}_{\boldsymbol{\theta}} \mathbf{u}, \tag{7}$$

where the approximation on the right is accurate when $\|\mathbf{u}\|$ is small. This relation is the basis for the proposed FIM estimator: we can estimate $D_\alpha$ using the FR runs statistic computed using the datasets from $p = p_{\boldsymbol{\theta}}$ and $q = p_{\boldsymbol{\theta}+\mathbf{u}}$ for various perturbations $\mathbf{u}$ on $\boldsymbol{\theta}$, then solve for $\mathbf{F}_{\boldsymbol{\theta}}$.

The above is the key relation that allows one to estimate $\mathbf{F}_{\boldsymbol{\theta}}$ directly from the data. In particular, assume that the experimenter can generate a data set $\mathbf{X}_{01}, \ldots \mathbf{X}_{0N_p}$ from the reference density $p_{\boldsymbol{\theta}}$ and $M$ other data sets $\mathbf{X}_{k1}, \ldots \mathbf{X}_{kN_q}$, $k = 1, \ldots, M$, from perturbed densities $p_{\boldsymbol{\theta}+\mathbf{u}_1}, \ldots, p_{\boldsymbol{\theta}+\mathbf{u}_M}$, where $\mathbf{u}_i \neq \mathbf{u}_j, i \neq j$. Furthermore, assume that the experimenter has full control of $\mathbf{u}$ and that $M$ is at least equal to $d(d+1)/2$, where $d$ is the dimension of the parameter space. Let $\mathbf{Q} = [Q_1, \ldots, Q_M]^T$ be the $M$ different values of $Q_{pq}$ that are computed from the FR runs statistic applied to the pairs of datasets $\{\mathbf{X}_{0i}\}_{i=1}^{N_p} \cup \{\mathbf{X}_{ki}\}_{i=1}^{N_q}$, $k = 1, \ldots, M$.

To solve for $\mathbf{F}_{\boldsymbol{\theta}}$ in (7), we vectorize $\mathbf{F}_{\boldsymbol{\theta}}$ by including the distinct lower triangular values of the FIM and convert the equation in (7) to a linear function of this quantity. A least squares estimator of $\mathbf{F}_{\boldsymbol{\theta}}$ can then be easily constructed. Define the $d(d+1)/2$ element vector of direction component pairs $[u_{11}^2, \ldots, u_{1d}^2, 2u_{11}u_{12}, \ldots, 2u_{1(d-1)}u_{1d}]$, and likewise for the other $M-1$ directions $\mathbf{u}_2, \ldots, \mathbf{u}_M$. Concatenate all of these direction component pair vectors in the $M \times d(d+1)/2$ matrix $\mathbf{U}$. Define the $d(d+1)/2$ element FIM vector:

$$\mathbf{F}_{\text{vec}} = [F_{11}, \ldots, F_{dd}, F_{12}, \ldots, F_{d(d-1)}]^T \tag{8}$$

This formulation allows us to write (7) as $\mathbf{Q} \approx \mathbf{U}\mathbf{F}_{\text{vec}}$ and to solve for $\mathbf{F}_{\text{vec}}$ using the least squares estimator given by:

$$\widehat{\mathbf{F}}_{\text{vec}} = (\mathbf{U}^T\mathbf{U})^{-1}\mathbf{U}^T\mathbf{Q}, \tag{9}$$

where $\mathbf{Q}$ is the $M \times 1$ vector defined above. This approach, however, does not ensure that the resulting estimate is positive semidefinite (PSD). We propose an alternate estimator that further constrains the problem by ensuring the solution is PSD.

For a given parameter, $\boldsymbol{\theta}$, an alternative estimator can be defined for estimating the corresponding FIM. Specifically, we can perturb the individual components of $\boldsymbol{\theta}$ independently to iteratively reconstruct the diagonal FIM values, $\widehat{F}_{ii}$, using a least squares procedure. This results in an estimate of the first $d$ components of $\mathbf{F}_{\text{vec}}$ and provides an additional set of constraints with which to further regularize the full FIM reconstruction. To estimate the full FIM, we propose the following semi-definite program:

$$\begin{aligned}
\underset{F_{vec}}{\text{minimize}} \quad & ||\mathbf{U}\mathbf{F}_{\text{vec}} - \mathbf{Q}||^2 \\
\text{subject to} \quad & \mathbf{F}_{\text{vec}}(i) = \widehat{F}_{ii},\ i \in 1\dots d, \\
& \text{mat}(\mathbf{F}_{\text{vec}}) \succeq 0,
\end{aligned} \quad (10)$$

where the mat() operator converts the vectorized FIM to a full matrix representation. This formulation aims to minimize the $\ell_2$ prediction error with respect to the vectorized FIM, subject to the constraint that the diagonal values are the same as those estimated individually and that the full FIM is PSD. This formulation is motivated by Boyd's constrained least squares covariance estimation algorithm in [13].

In principle the pertubation parameter, $\mathbf{u}$, could be selected according to an optimal design criterion, e.g., minimize an empirical estimate of the FIM estimation error. However, any set of $\mathbf{u}_i$'s that make the matrix $\mathbf{U}$ of rank $d(d+1)/2$ will lead to an asymptotically consistent FIM estimator as $N_p, N_q \to \infty$ and $||\mathbf{u}_i|| \to 0$, $i = 1,\dots, M$. In the sequel, we select the perturbations randomly over a small radius ball in $\mathbb{R}^d$.

## 4 Results

In this section we validate the proposed procedure. In the first experiment, we estimate the mean square error (MSE), as measured by the squared Frobenius norm of the estimator error, of the FIM estimator for different data dimensions. In the second experiment, we empirically analyze a complicated signal processing chain for which it becomes impossible to derive an analytical relation between parameters and observations. We analyze a cochlear implant (CI) simulator similar to the one in [14]. Through the relationship between the CRLB and the FIM, we empirically estimate the minimum variance of the best unbiased estimator of the signal energy at different frequency components in the input speech signal (parameters, $\boldsymbol{\theta}$), given only current levels of the different electrodes. This is the problem that CI patients must solve - given electrode stimulation, resolve the input speech. Here we assume an equal number of samples from $p$ and $q$, therefore $\alpha = 1/2$ implicitly.

**Empirical FIM Estimation:** The objective of this experiment is to compare estimates of the FIM using the method proposed here with the closed form solution for Gaussian data. Let us consider a Gaussian distribution parameterized by the mean $\boldsymbol{\mu}(\boldsymbol{\theta}) = \boldsymbol{\theta}$ and unity covariance. We aim to estimate the FIM at $\boldsymbol{\theta} = \mathbf{0}$, $\mathbf{F_0}$, for different data dimension.

Let us denote a $K$-dimensional Gaussian PDF with mean $\mathbf{0}_{K\times 1}$ and covariance $\sigma^2 \mathbf{I}_{K\times K}$ by $p_{\mathbf{0},\sigma^2\mathbf{I}}(\mathbf{x})$, where $\mathbf{I}$ is the identity matrix and $\mathbf{0}$ is the zero vector. We perturb this distribution by $\mathbf{u}$ about $\boldsymbol{\theta} = \mathbf{0}$ and denote the resulting PDF by $p_{\mathbf{u},\sigma^2\mathbf{I}}(\mathbf{x})$, where $\mathbf{u} \sim \mathcal{N}(0, \sigma_{\mathbf{u}}^2)$. For this experiment, the closed form estimate of the FIM is known (given by the identity matrix, $\mathbf{I}_{K\times K}$, since $\sigma^2 = 1$). It is important to note that for this scenario, $d = K$ since the distribution is parameterized by the mean. We estimate the FIM from empirical data using the algorithm proposed here and evaluate the FIM MSE. In particular, we draw $N = 1000$ points from both $p_{\mathbf{0}, \sigma^2 \mathbf{I}}(\mathbf{x})$ and $p_{\mathbf{u}, \sigma^2 \mathbf{I}}(\mathbf{x})$ and estimate the $f$-divergence for multiple values of $\mathbf{u}$ ($\sigma_{\mathbf{u}}^2 = 0.05$) for different values of $K$ ranging from 4 to 14. Note that this corresponds to $\alpha = 1/2$ and $N_p = N_q = N$ in (3) and (4). For each dimension, we perform $M = 50K(K+1)/2$ random perturbations. From this set of parameters, we use the least squares estimator in 3 to obtain an estimate of the FIM.

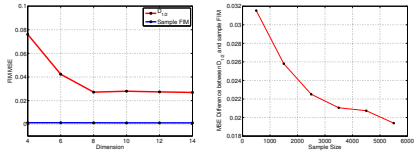

Figure 1: (a) MSE of the FIM estimate for the proposed $D_{\frac{1}{2}}$-based estimator (b) The MSE difference between the two estimators in (a). The benchmark is an oracle estimator that knows that the distribution is multivariate Gaussian.

For this problem the FIM is $\sigma^{-2} \mathbf{I}_{d \times d}$. In Fig. 4 (a), we plot the MSE between the proposed estimator of the FIM and the true FIM as a function of dimension. The MSE estimate shown in Fig. 1 is the result of averaging over 25 Monte Carlo simulations and for every run we use a different realization of $\mathbf{u}$. Because we use a large value of $M$ (50 times more samples than the minimum required), the estimate is stable from iteration to iteration. This allows us to use a smaller number of Monte Carlo iterations for estimating the MSE. The benchmark against which we compare is an oracle estimate of the FIM that assumes that we know that the distribution is multivariate Guassian but do not know its covariance: we estimate our benchmark FIM by inverting the sample covariance estimator and refer to it as the "sample FIM estimate". We see that the error between the two estimators is comparable, though the sample FIM has a lower MSE given the additional information. Interestingly, this error decreases for higher dimensions since, for the same level of perturbation, it becomes easier to reliably estimate $D_{\frac{1}{2}}(p, q)$ from data.

In Fig. 4 we plot the MSE difference between the $D_{\frac{1}{2}}$-based estimator and the sample estimator as the sample size increases. We fix the dimension to $K = 20$ and use the same perturbation variance, $\sigma_{\mathbf{u}}^2 = 0.05$. As expected, the sample FIM starts with a lower MSE due its access to the additional side information that the distribution is Gaussian, however the difference between the two estimators decreases as the sample size increases. The proposed estimator performs almost as well as the oracle estimator even though it does not have access to this side information.

**Estimating the CRLB in a Complex System:** Beyond estimating the Fisher Information itself, we can use the non-parametric divergence to characterize the difficulty of an estimation problem. It is well known that the FIM can be used to benchmark the performance of an estimator (via the CRLB). If $\widehat{\boldsymbol{\theta}}$ is an unbiased estimator of $\boldsymbol{\theta}$, then the inverse FIM lower bounds the covariance matrix of the estimator by $\text{cov}(\widehat{\boldsymbol{\theta}}) \geq \mathbf{F}_{\boldsymbol{\theta}}^{-1}$. The CRLB is known in closed form for certain distributions; however, for complex signal processing systems, it becomes impossible to characterize the CRLB for a specific parameter unless simplifying assumptions are made. We consider a signal processing simulator for a 16-electrode cochlear implant device. CI's provide hearing in patients with damage to sensory hair cells by stimulating the 16 electrodes with appropriate energy. In a CI, a microphone picks up audio from the environment, a speech processor decomposes the speech in sub-bands and identifies the appropriate amount of current with which to stimulate each of the electrodes, and stimulates the electrode array. We have implemented a CI simulator in Matlab, similar to the Simulink simulator by Loizou in [14] (see Fig. 2). This speech processor includes linear (filtering) and non-linear (rectification, intensity compression) operations. The input of the processor is the spectrum of the speech signal ($\boldsymbol{\theta}$) and the output of the simulator ($\mathbf{x}$) represents the current level (in ampères) measured at the 16 output electrodes. Here, we are interested in characterizing the bound on estimator variance for the amplitude spectrum of the speech signal when we observe the electrode current level.

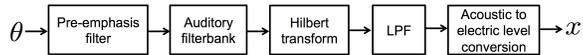

Figure 2: The cochlear implant signal processing chain.

We evaluate the algorithm on the TIMIT corpus - a phonemically balanced database that spans multiple speakers and multiple regional dialects in the US [15]. We sample 100 sentences from this database, resample to 8 kHz, normalize the average energy to a fixed value, and estimate the FIM as in Sec. 3. We first process each speech signal through the CI simulator. The simulator analyzes each input speech signal using a 10ms frame size (with 80% overlap) and estimates the current level at each of the 16 electrodes per frame. If the total number of frames is $N$, this results in a data matrix of $N \times 16$. We denote this matrix by $\mathbf{X}_{\boldsymbol{\theta}}$. For a given frequency component, we perturb this parameter by a small value $\mathbf{u}$ by simply adding a sinusoid of appropriate energy to the original speech. We then process the resulting audio through the CI simulator to generate a new data matrix, $\mathbf{X}_{\boldsymbol{\theta}+\mathbf{u}}$.

In this simulation, rather than estimating the FIM in a single step, we restrict ourselves to the diagonal components of the FIM. We sweep through the frequency spectrum by iteratively perturbing the spectrum of the original speech signal and generating a new data set for each perturbation (through the CI simulator) then iteratively estimate the diagonal components of this matrix one at a time. For every speech signal, at every frequency we perform $M = 10$ perturbations with $\mathbf{u}$ drawn from a normal distribution of mean 3.5 and variance 1.

In Fig. 3 (top), we plot the average divergence as a function of frequency from 150 Hz to 3.6 kHz. As the figure shows, there is a general increasing trend in the plot. This makes sense intuitively as the average speech spectrum contains more signal energy at low frequencies. The 16 peaks and troughs of the divergence measure are also consistent with the filter bank used in the simulator. In Fig. 3 (bottom), we plot the square root of the estimated CRLB over the average frequency spectrum. As the figure shows, the amplitude and the standard deviation (SD) of the estimator are often of the same order. Indeed, for a number of frequencies, the SD is greater than the mean indicating that, at those frequencies, even the best estimator will have difficulty identifying significant differences between the true signal energy and no energy. This is consistent with the poor fidelity speech in real CIs [16].

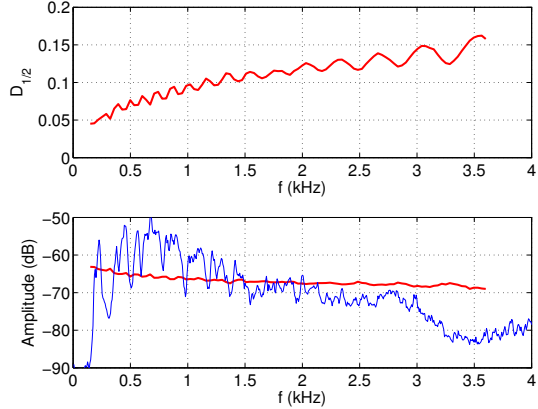

Figure 3: (Top) $D_{\frac{1}{2}}$ as a function of frequency (Bottom) The average speech spectrum and the empirically-estimated CRLB

In addition to calculating the diagonal CRLB for the TIMIT database, we also compare the difficulty of estimating specific vowels by calculating the full CRLB matrix for each vowel. We use the vowels from male speaker "m01" in the Hillenbrand data set [17]. As before, we resample the data to 8 kHz, normalize the average energy to a fixed value, and perform the experiment as in Sec. 3. To manage computational complexity, rather than estimating the full FIM associated with all input frequencies, we reconstruct the $16 \times 16$ sub-matrix associated with the center frequencies of the auditory filter bank in Fig. 2. We justify this by noting that the CI model is most sensitive at these center frequencies (see peaks in Fig. 3 (top)); this sub-matrix acts as a best-case estimation bound. For each vowel, every component of **u** is drawn from a normal distribution of mean 3.5, variance 1 for and a total of $M = 10d(d+1)/2 = 1360$ perturbations per FIM estimate.

We estimate $\mathbf{F}_{\boldsymbol{\theta}}$ using the $D_{1/2}$-based estimator in Sec. 3 and we invert to calculate the CRLB, $\mathbf{C}_{\boldsymbol{\theta}} = \mathbf{F}_{\boldsymbol{\theta}}^{-1}$. Prior to inverting, we use the Bayesian-based diagonal loading procedure by Haff for regularization [18]. For each vowel, we compute the volume of the CRLB matrix, weighted by a perceptual weight-ing filter giving more weight to bands that contain more signal energy. The weighting matrix, $\mathbf{W}$, is a diagonal matrix with the components given by the signal energy at the location of the 16 frequencies associated with $\boldsymbol{\theta}$. Perceptual weighting of frequency bands by signal energy is common in many speech applications [19, 20]. The volume of the weighted CRLB matrix is given by $\text{Vol} = \log(\det(\mathbf{VDWV}^\text{T}))$, where $\mathbf{C}_{\boldsymbol{\theta}} = \mathbf{VDV}^\text{T}$ is the SVD of $\mathbf{C}_{\boldsymbol{\theta}}$. This serves as a proxy for the uncertainty associated with estimating the spectrum of the speech signal from only the current at the electrodes. It is expected that vowels with a small volume will be more intelligible than vowels with a large volume by CI patients. In Fig. 4 we plot the CRLB volume for the vowels in the study.

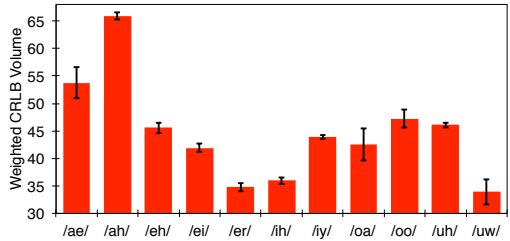

Figure 4: The CRLB volume for each of the 11 vowels in [17].

A similar behavioral study was conducted by Dorman et. al in [21]. Using the same dataset, the authors encode the vowels with a simulated CI and conduct an intelligibility assessment task where they ask participants to correctly identify the encoded vowels. As in Fig. 4, Dorman shows that there are two vowels that prove to be difficult for the listeners, /ah/ as in "hod" (43% correct classification rate) and /ae/ as in "had" (68% correct classification rate) (Table II in [21]). Overall, correlating the average CRLB volumes in Fig. 4 with the intelligibility values reveals a correlation coefficient of -0.801.

## 5  Conclusion

In this paper we have proposed a new data-driven approach for estimating the Fisher information by exploiting the relationship between the FIM and the family of $f$-divergence measures. The estimator relies on multiple perturbations of a signal model followed by a semidefinite program that reconstructs the FIM. We test the algorithm by evaluating its ability to reconstruct the FIM for a signal model with analytical ground truth FIM. In addition, we showed that we can predict the difficulty of resolving different speech samples processed through a CI by comparing intelligibility values estimated through behavioral experiments with the volume of the CRLB. Future work will analyze the asymptotic properties of this estimator and discretization bias. This would allow us to better understand the approximation error as a function of the underlying density and could serve as a starting point for developing theory for selecting parameter perturbations.